\documentstyle[aps,preprint,epsfig,psfrag]{revtex}
\newcommand{\be}{\begin{equation}} 
\newcommand{\ee}{\end{equation}}
\newcommand{\ba}{\begin{eqnarray}}
\newcommand{\ea}{\end{eqnarray}}
\newcommand{\no}{\nonumber \\}
\begin{document}
\begin{titlepage}
\pagestyle{empty}
\vspace{1.0in}
\begin{flushright}
May 2000
\end{flushright}
\vspace{.1in}
\begin{center}
\begin{large}
{\bf Moving Domain Walls in $AdS_5$ and Graceful Exit from Inflation}
\end{large}
\vskip 0.5in
Chanyong Park\footnote{chanyong@hepth.hanyang.ac.kr} 
 and Sang-Jin Sin\footnote{ sjs@hepth.hanyang.ac.kr} \\
\vskip 0.2in
{\small {\it Department of Physics, Hanyang University, Seoul, 133-791, Korea
}}
\end{center}
\vspace{1 cm}
\tightenlines 

\begin{abstract}
We consider moving-brane-solutions in AdS type back ground. In the 
first  Randall-Sundrum configuration, there are  two branes at  
fixed points of the orbifold symmetry. We point out that if 
one brane is fixed and the other brane is moving, the 
configuration is still a solution provided  the moving brane 
has a specific velocity determined by its tension and bulk cosmological constant. 
In the absence of the $\bf Z_2$ symmetry, we can construct multi-brane 
configurations by patching AdS-Schwarzshild solutions.  In this 
case, we show that the 4-dimensional effective  
cosmological constant on the brane world is not well defined.  
We find a condition for a brane to be stationary.   
Using the brane scattering, we suggest a scenario of inflation on the brane universe
during a finite time, i.e, a scenario  of a graceful exit of inflation. 
    
\end{abstract}

\vspace{.3in}\noindent PACS numbers: 12.10.-g, 11.10.Kk, 11.25.Mj 
\end{titlepage}      
\tightenlines

\section{Introduction}
Recently, Randall and Sundrum made  an proposal that we may live in a four dimensional section of 
5 dimensional universe \cite{RS}. The key point of the first one is that 
gravitational warp factor can generate a small number to solve the 
hierarchy problem. The idea of the second one is that due to the  the attraction of the 
brane energy, the metric fluctuation around the domain wall admits a bound state
and the background cosmological constant makes this 'bound state' a zero-mode to be identified as a 
graviton. The brane tension is fine tuned so that 
the effective cosmological constant on the domain wall is zero and the brane is stationary.   
If the fine tuning is relaxed, bulk metric depends
on time and the brane inflates\cite{kaloper}. These solutions can 
be used to discuss the cosmology of the RS-model\cite{kaloper,rscos}.

More recently, Kraus pointed out that these solutions can be interpreted 
as motion of a domain wall in a stationary background 
\cite{Kraus}.\footnote{
Ref.\cite{kiritsis} discussed the motion of a 'probe brane' in fixed background  
using the Dirac-Born-Infeld action.}
The idea of \cite{Kraus} is that for the given cosmological 
constant we can glue two pieces of AdS-Schwartzshild (AdSS) solutions 
along the moving domain wall using the Israel matching conditions 
to construct the new space-time containing the domain wall. 
The matching condition determines the motion of the domain wall in terms of
the domain wall tension and the bulk cosmological constant.
   
The moving brane solutions are interesting since those can lead 
us scenarios for the graceful exit from the inflation in the Randall-Sundrum type 
cosmology. 
For example, suppose 
we live in a brane  whose brane tension is fine tuned to have flat 
4 dimensional geometry. Now if there is a moving brane with finite thickness, the  collision of  
our brane with such a moving brane provide inflation in our brane world for the  finte collision 
time, as we will argue.
Having explicit analytic solution for the two moving branes with 'finite thickness' is not
an easy task. So finding solutions for two moving thin branes in a fixed background is the first step 
in this direction and this is one of goals of this paper. 
Generalization of the RS1 model to time dependent 
case has been discussed in \cite{kaloper} wherein either the distant between the two branes
is fixed or the  bulk metric itself has  explicit time dependence.

We first examine the case with orbifold symmetry, which is the generalization of the 
Randall-Sundrum to the moving brane case.  We try to solve the full 
5 dimensional bulk Einstein solution with the warp factor corresponding to the moving branes. 
It turns out that the Einstein equation dictates that the 
information of the brane location must disappear from the bulk 
metric. However the motion of the brane can not be  arbitrary:
it is fixed by the tension of the brane and the bulk cosmological constant.
We will get a solution with two branes; one is stationary and the other is 
moving. In the presence of the reflection symmetry, the tensions of 
the two branes must have opposite sign. 
To get the solution where both of the branes have  positive tension, we have to 
abandon the orbifold symmetry. We use Kraus' method. We show that the 4-dimensional effective  
cosmological constant on the brane world in the absence of the 
reflection symmetry, is not well defined.  
We find a condition for a brane to be stationary.   
Finally we consider the brane scattering using these solutions and 
suggest a scenario for the inflation during finite duration\cite{linde}.

\section{Summary of RS-model}

We start with following action:
\be
S = S_{gravity} + \Sigma_i S_{i({\rm wall})} ,  \label{action}
\ee
where  $S_{gravity}$ is given by
\be
S_{gravity} = \int d^5 x \sqrt{g} \left[ \frac{1}{2\kappa^2} {\cal R}
            - \Lambda  \right],
\ee
with $\Lambda$ being a cosmological constant and $g_{\mu\nu}$ a
metric of five-dimensional space-time.
The domain wall action can be written as
\be
S_{i(wall)} = - \int d^5 x \sqrt{g} {\cal L}_{i(wall)} \delta(y - R_i (t)) ,
\ee 
where $y$ is a coordinate of a transverse direction. Here the 
delta function implies that the domain wall lies at the position 
$y=R_i (t)$ and can possibly move. 

Assuming that all excitation modes of the matter on the wall are absent, 
the action Eq. (\ref{action}) is reduced to the 
\be
S = \int d^5 x \sqrt{g} \left[ \frac{1}{2\kappa^2} {\cal R} 
    - \Lambda \right] - \sum_i \sigma_i \int d^5 x \sqrt{g} 
     \delta(y - R_i (t)) ,  \label{sact}
\ee
where $\sigma_i$ is a tension of the $i$-th domain wall. 
Since we are interest to the AdS space-time in the
bulk, only the case $\Lambda < 0$ will be considered.
From Eq. (\ref{sact}), Einstein equations become
\be
{{\cal R}^{\mu}}_{\nu} - \frac{1}{2} {{\delta}^{\mu}}_{\nu} {\cal R}
     = - \kappa^2 {T^{\mu}}_{\nu} .
\ee
Here, the energy-momentum tensor ${T^{\mu}}_{\nu}$ is given by
\ba
&& {T^{\mu}}_{\nu} = - |\Lambda| {\rm diag} (1, 1, 1, 1, 1)  
                    + \Sigma_i {{T_{i(wall)}}^{\mu}}_{\nu}, \no
&& {{T_{i(wall)}}^{\mu}}_{\nu} = \sigma_i {\rm diag} (1, 1, 1, 1, 0) 
                                \delta(y - R_i (t)) .
\ea
where ${{T_{i(wall)}}^{\mu}}_{\nu}$ is an
energy-momentum tensor on the i-th wall.

The vacuum solution of the Einstein equation in the presence of 
the negative cosmological constant $\Lambda$ is AdS space:
The Randall-Sundrum model is constructed by joining two AdS  
\be
ds^2 = e^{2ky} (-dt^2 + \delta_{ij} dx^i dx^j ) +dy^2 , \label{adsf}
\ee 
where 
 \be
k^2 = \frac{\kappa^2}{6} |\Lambda| , \label{brel}
\ee

The Randall-Sundrum\cite{RS} solution  is 
that domain wall solution can be obtained by 
joining two 'inner-part' of the AdS spaces along the hyper
surface at $y=0$   
\cite{verlinde}:
\be
ds^2 = e^{-2k\mid y\mid} (-dt^2 + \delta_{ij} dx^i dx^j ) +dy^2 . \label{adsf}
\ee 
The solution has manifest $\bf Z_2$ summetry $y\to -y$.
The domain wall at $y=0$ has tension given by
\be
\sigma=6k/\kappa^2.
\ee

Now, for later purpose, let's ask what should be the solution if we 
put the domain wall along $y=R$.
Due to the  ${\bf Z_2}$ symmetry,  the metric should be as 
follows:
\ba
ds^2 &=& e^{-2k \mid y-R \mid } (-dt^2 + 
         \delta_{ij} dx^i dx^j ) +dy^2
         \;\;\;  {\rm for} \;\;\; \hbox{near~~~  y=R}, \no
     &=& e^{-2k \mid y+R \mid } (-dt^2 + \delta_{ij} dx^i dx^j ) +dy^2 
         \;\;\; {\rm for} \;\;\; \hbox{near~~~  y=-R}, \label{rsmet1}
\ea The sign in the warp factors is chosen such that as $\mid 
y\mid \to \infty$ we are going deep inside the AdS spaces. 
Physically, there is no difference 
whether we have domain wall at $y=0$ or $y=R$ apart from the over 
all scale $e^{2kR}$. This can be verified by observing that
above metric can be written as 
\be
ds^2 = e^{-2{k}|y|+2k R } (-dt^2 + 
         \delta_{ij} dx^i dx^j ) +dy^2  , \hbox{for } |y| \ge R. \label{met2w} 
\ee
Notice that $y=R$ and $y=-R$ are identified and 
there is no domain wall at $y=0$ since the region $-R \le y \le R$ is not 
in the universe in this construction. This is the 1-wall solution with positive tension (RS2)  \cite{RS}.  

However, in the region $-R \le y \le R$,  the same expression (\ref{rsmet1}) can be written as   
\be
ds^2 = e^{2{k}|y|-2k R } (-dt^2 + 
         \delta_{ij} dx^i dx^j ) +dy^2. \label{met3w} 
\ee
This means that if $k$ is positive, there is a domain wall with negative tension at $y=0$,
as well as positive tension domain wall at  $y=R$. Therefore (\ref{met3w}) represent 
solution with two domain walls (RS1).
Therefore, according to the  region we are looking at,  the solution 
(\ref{rsmet1}) can gives either RS2 or RS1.
The RS solution is not a geometry for one ads space with a 
brane but a patching two ads along a brane\cite{verlinde}.
See figure 1(a,b). Notice that the figure 1.(b) also tells us that as $R \to \infty$, RS1 is reduced 
to RS2. 
Later on, we will see that replacing $R\to R(t)$ is still solution for
specific $R(t)$. 

\vspace{-3cm}
\begin{figure}
  \unitlength 1mm
   \begin{center}
      \begin{picture}(25,100)
      %\graphpaper[2](-60,0)(190,80)
      \put(-70,30){\epsfig{file=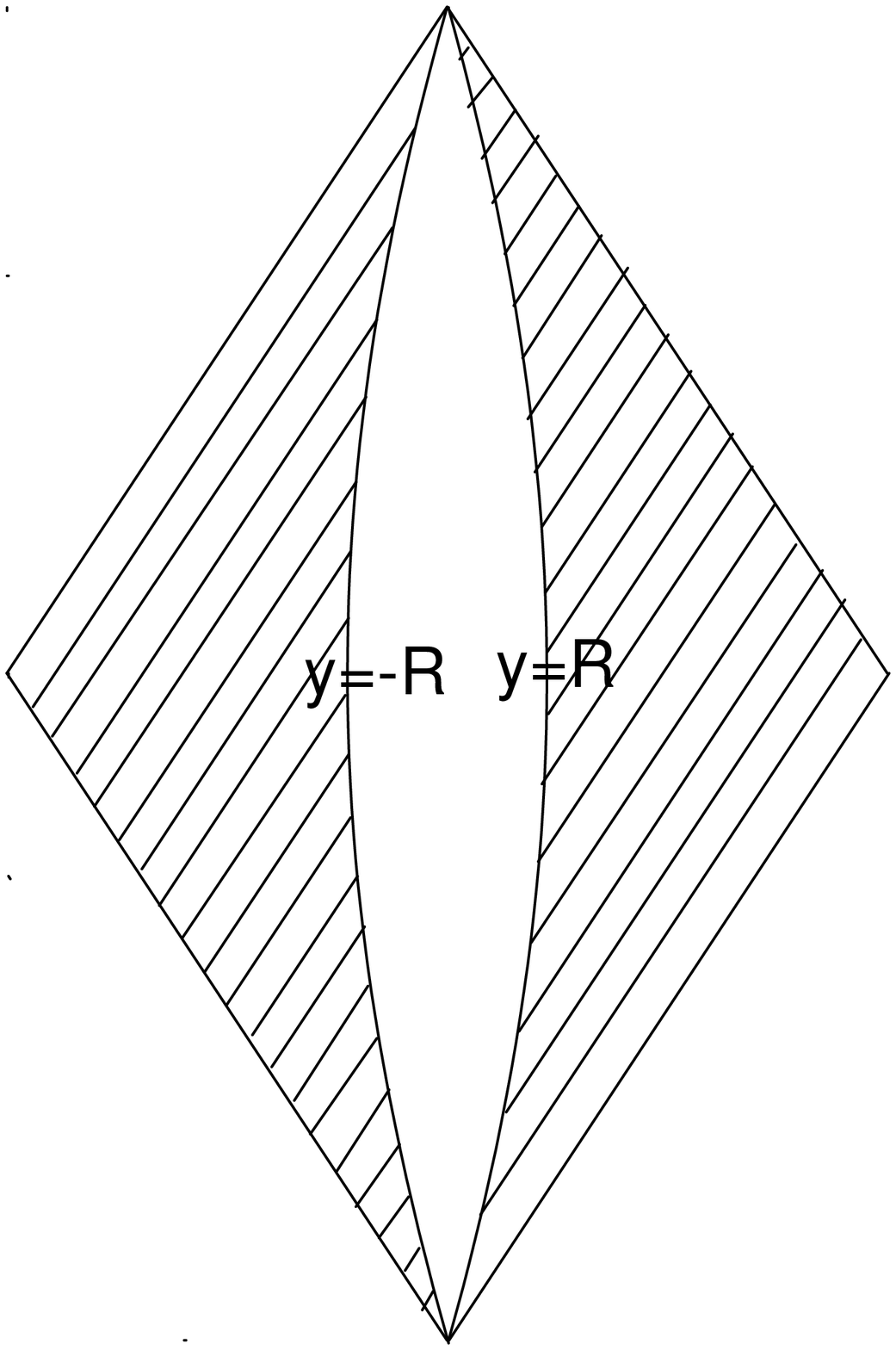,width=8cm,height=8cm}}
      \put(0,30){\epsfig{file=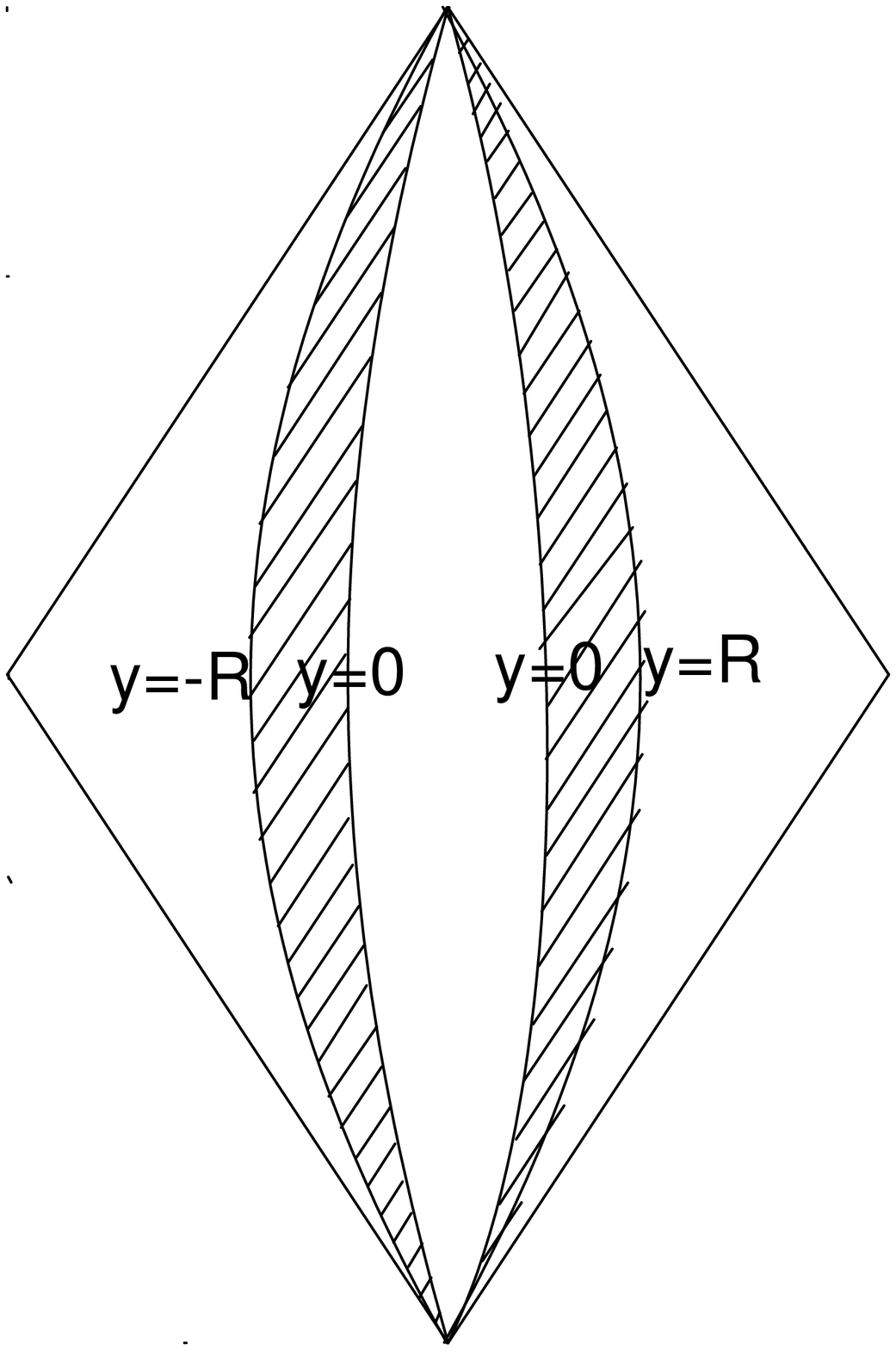,width=8cm,height=8cm}}
      \put(-30,30){$(a)$ }
      \put(40,30){$(b)$ }
      \end{picture}
   \end{center}
\vspace{-3cm}
\caption{(a) The Penrose diagram for RS2 solution with 1-brane:
The brane is located at $y=R$ instead of $y=0$. 
(b) The same for RS1 with two branes.
 Replacing $R\to R(t)$ is still solution for specific $R(t)$.}
\label{fig1}
\end{figure}

%\newpage

If a constant $R$ is replaced with time dependent $R(t)$, we would be  
considering  a moving domain wall. 
In the next section we are looking for a solution which has a warp factor 
corresponding  moving domain walls.  

\section{Moving Domain Walls}

\subsection{ RS1-type model}
In this section, we look for a solution with moving domain walls
located at $y=R(t)$ and $y=0$. If we impose ${\bf Z_2}$
symmetry, there must be a mirror image of the moving wall at
$y=-R(t)$. So we are looking for the moving version of the solution (\ref{rsmet1}). 
Therefore we start from the
following ansatz for the region $-R(t) \le y \le R(t)$:
\be
ds^2 = e^{-2k |y| + 2k R(t)} (-dt^2 + H(t,y)^2 \delta_{ij} dx^i dx^j ) 
       + b(t,y)^2 dy^2 ,  \label{ans}
\ee
where $k$ in Eq. (\ref{met3w}) is replaced with $-k$ and 
this metric is RS1-type metric which has two walls at $y=0$
and $y=R(t)$. Since we impose ${\bf Z_2}$ symmetry, 
the above metric must be invariant under $y \to -y$. 
To determine $H(t,y)$ and $b(t,y)$, 
we have to solve five-dimensional Einstein equations. 
If we write the metric as 
\be
{ds_5}^2 = -n(t,y)^2 dt^2 + a(t,y)^2 \delta_{ij} dx^i dx^j 
             + b(t,y)^2 dy^2.
\ee 
Einstein equations are given by \cite{kaloper}
\ba
G_{00} &=& 3 \left[ \frac{\dot{a}}{a} \left( \frac{\dot{a}}{a} +
     \frac{\dot{b}}{b} \right) - \frac{n^2}{b^2} \left\{ 
     \frac{a^{\prime\prime}}{a} + \frac{a^{\prime}}{a} \left(
     \frac{a^{\prime}}{a} - \frac{b^{\prime}}{b} \right) \right\}
     \right] = \kappa^2 T_{00} , \no   
G_{ii} &=& \frac{a^2}{b^2} \left[ \frac{a^{\prime}}{a} \left( 
     \frac{a^{\prime}}{a} +2 \frac{n^{\prime}}{n} \right) 
     -\frac{b^{\prime}}{b}
     \left( \frac{n^{\prime}}{n} +2 \frac{a^{\prime}}{a} \right) +2
     \frac{a^{\prime\prime}}{a} + \frac{n^{\prime\prime}}{n} \right] \no
       &&+ \frac{a^2}{n^2} \left[ \frac{\dot{a}}{a} \left( -
     \frac{\dot{a}}{a} +2 \frac{\dot{n}}{n} \right) + \frac{\dot{b}}{b}
     \left( -2 \frac{\dot{a}}{a} + \frac{\dot{n}}{n} \right) -2
     \frac{\ddot{a}}{a} - \frac{\ddot{b}}{b} \right] 
     = \kappa^2 T_{ii} , \no
G_{55} &=& 3 \left[ \frac{a^{\prime}}{a} \left( 
     \frac{a^{\prime}}{a} + \frac{n^{\prime}}{n} \right) 
     -\frac{b^2}{n^2} \left\{ \frac{\dot{a}}{a} \left( 
     \frac{\dot{a}}{a} - \frac{\dot{n}}{n} \right)
     \frac{\ddot{a}}{a} \right\} \right] = \kappa^2 T_{55}, \no
G_{05} &=& 3 \left[ \frac{n^{\prime}}{n} \frac{\dot{a}}{a} 
     + \frac{a^{\prime}}{a} \frac{\dot{b}}{b} 
     - \frac{\dot{a}^{\prime}}{a} \right] = 0 ,  \label{eineq}
\ea
where we use dot and prime to describe a derivative with respect
to $t$ and $y$ respectively.
Due to ${\bf Z_2}$ symmetry, 
it is sufficient that we pay attention to the right-hand side of the domain wall only.
Our ansatz means that, 
$n(t,y)$ and $a(t,y)$ are given by
\ba
n(t,y) &=& e^{-ky +kR(t)}, \no
a(t,y) &=& n(t,y) H(t,y) = e^{-ky + kR(t)} H(t,y) . 
           \label{ans}
\ea
in the region $0 \le y \le R(t)$.
Using these equations, $G_{05}$ in the bulk reads
\be
\left[ \left( \frac{H^{\prime}}{H} \right) - k \right] 
   \frac{\dot{b}}{b}  
      = \left( \frac{{\dot{H}}^{\prime}}{H} \right) + k \dot{R} 
        \left( \frac{H^{\prime}}{H} \right) . \label{g05}
\ee   
Now we take the gauge in which $b(y,t)=1$. 
Then  Eq. (\ref{g05}) can be solved by
\be
H(t,y) = e^{- k R(t)} G(y) . \label{fh}
\ee  
for any $G(y)$.
Inserting Eq. (\ref{fh}) into Einstein equations,
the other Einstein equations can be rewritten as
\ba
G_{00} &=& -3 n^2 \left[ \frac{G^{\prime\prime}}{G} 
           + \left( \frac{G^{\prime}}{G} \right)^2 
           - 4k \frac{G^{\prime}}{G} + 2k^2 \right] 
           = - n^2 \kappa^2 |\Lambda| , \no
G_{ii} &=& n^2 H^2 \left[ 2 \frac{G^{\prime\prime}}{G}
           + \left( \frac{G^{\prime}}{G} \right)^2
           - 8k \frac{G^{\prime}}{G} + 6k^2 \right]
           = n^2 H^2 \kappa^2 |\Lambda| , \no 
G_{55} &=& 3 \left[ \left( \frac{G^{\prime}}{G} \right)^2 
           - 3k \frac{G^{\prime}}{G} + 2k^2 \right] 
           = \kappa^2 |\Lambda| . \label{exeq}
\ea
%Note that the wall does not give any effects to the bulk Einstein equations. 
If we set $G(y)=e^{mky}$, 
only the case of $m=0$ satisfies all equations in Eq. (\ref{exeq}) 
with following relation: 
\be
k^2 = \kappa^2 |\Lambda| /6 . \label{relb}
\ee
Finally, the metric in the region $0 \le y \le R(t)$ is 
\be
ds^2 = e^{-2ky + 2kR(t)} (-dt^2 + e^{- 2kR(t)} 
       \delta_{ij} dx^i dx^j) + dy^2  . \label{posmet}
\ee 
Due to ${\bf Z_2}$ symmetry 
the metric in the region $-R(t) \le y \le 0$ is
\be
ds^2 = e^{ 2ky + 2kR(t)} (-dt^2 + e^{- 2kR(t)} 
       \delta_{ij} dx^i dx^j) + dy^2. 
\ee
Therefore the metric in $-R(t) \le y \le R(t)$ can be written 
simply as  
\be
ds^2 = e^{- 2k |y|} (- e^{2k R(t)}dt^2 
       + \delta_{ij} dx^i dx^j) + dy^2 \label{sol2}, 
\ee
which is a AdS metric in the bulk by redefining the time. 
At $y=0$, there is a static wall. 
By integrating the Einstein equation across the static wall domain wall, 
we get the relation of  the tension and $k$: 
\be
\sigma= \frac{6 k}{\kappa^2}: = \sigma_c  \label{jump}.
\ee 
This, together with Eq. (\ref{relb}), gives
the exact value of the tension of  the static wall  in terms of the bulk cosmological 
constant: 
\be
\sigma = \pm \sqrt{\frac{6 |\Lambda|}{\kappa^2}}.
\ee
where $\pm=k/|k|$.

Note that the bulk Einstein equations do not determine $R(t)$, the position of 
the moving wall: for arbitrary function $R(t)$, (\ref{sol2}) is a solution of the 
bulk Einstein equations.  
As is well known,
$R(t)$ is determined by the Israel junction 
equation\cite{Israel,Kraus}. In terms of the  extrinsic curvature of the wall world-volume 
\be 
K_{\mu\nu} = \nabla_{\mu} n_{\nu}, \label{israel}
\ee
with $n_{\nu}$ the unit normal vector on the domain wall 
world-volume, the Junction condition reads
\be
\Delta K_{mn} = - \kappa^2 \left( T_{mn} - \frac{1}{3} {T^l}_l g_{mn}
                  \right),
\ee
where 
$\Delta K_{mn} := \lim_{y \to +0} K_{mn} - \lim_{y \to -0} =
{K^+}_{mn} - {K^-}_{mn}$. Here $g_{mn}$ is a metric on the wall and
$T_{mn}$ is a energy-momentum tensor of the wall. 
If we impose ${\bf Z_2}$ symmetry, $\Delta K_{mn}$ is equal to $2
{K^+}_{mn}$ due to ${K^+}_{mn} = - {K^-}_{mn}$.  
For `extremal' wall, the case where $T_{mn} = \sigma g_{mn}$, the
Israel junction equations become
\be
2{K^+}_{mn} =  \frac{\kappa^2}{3} \sigma g_{mn} .
\ee
To be explicit, let's write the metric as
\be
ds^2 = - d\tau^2 = - f(t,y) dt^2 + g(y) \delta_{ij} dx^i dx^j + dy^2 ,
\ee
and  let $u^{\mu}$ be the tangent velocity vector of the wall satisfying
$u^{\mu} u_{\mu} = -1$. Because the wall moves in $y$ direction,
$u^{\mu}$ is given by 
\be
u^{\mu} = (\sqrt{\frac{1+(\partial_{\tau} y)^2}{f}}, 0, 0, 0, 
           \partial_{\tau} y).
\ee
Since the unit normal vector $n^{\mu}$ satisfies $n^{\mu} u_{\mu} 
=0$, 
\be
n^{\mu} = (-\frac{\partial_{\tau} y}{\sqrt{f}}, 0, 0, 0, 
           - \sqrt{1+(\partial_{\tau} y)^2}) .
\ee 
Insert $n_{\mu}$ into Eq. (\ref{israel}), 
we obtain two junction equations: one is 
\be
%2 {K^+}_{00} &=& 2 \sqrt{f} \dot{\tau} {\partial_{\tau}}^2 y 
%                + \sqrt{1+(\partial_{\tau} y)^2} \partial_y f 
%              = - \frac{\kappa^2}{3} \sigma f , \no
2 {K^+}_{ii} = - \sqrt{1+(\partial_{\tau} y)^2} \partial_y g
            = \frac{\kappa^2}{3} \sigma g ,  \label{ije}
\ee   
and the other  $K_{00}$ equation does not give an independent one.
In our case, 
$f(t,y)$ and $g(t,y)$ are given by 
$f(t,y) = e^{ 2ky + 2k R(t)}$ 
and $g(t,y) = e^{ 2ky}$. 
The $ii$ component of Israel junction equations is reduced to 
\ba
  \sqrt{1+(\partial_{\tau} R(t))^2} 
             = - \frac{\sigma}{\sigma_c}.
\ea
Notice that if $k$ is positive, the tension of the moving wall must be
negative and
$R(t)$ is determined as 
\be
R(t) = v \tau,
\ee      
where $v = \pm \sqrt{|\sigma /\sigma_c |^2 -1}$.
Using $u^0 = dt/d\tau = \sqrt{\frac{1+(\partial_{\tau} y)^2}{f}}$,
\be
\tau(t) = \frac{1}{\sqrt{1+v^2}} t.
\ee
So $R(t)$ is given by
\be
R(t) = \frac{v}{\sqrt{1+v^2}} t .
\ee
%Since $R(\tau)$ depends on $\tau$ linearly
%${\partial_{\tau}}^2 R(\tau)=0$, hence
%the junction equation for ${K^+}_{00}$ is consistent with junction
%equations for ${K^+}_{ii}$.

From the bulk metric, the four-dimensional metric on the static wall at $y=0$ 
can be reduced to
\be
{ds_4}^2 = - d\tilde{t}^2 + \delta_{ij} dx^i dx^j ,
\ee
where $d\tilde{t} =e^{kR(t)}dt$. 
This metric is a Mikowskian one as is the Randall-Sundrum case. 
The metric on the moving wall at $y=R(t)$ is
\be
{ds_4}^2 = - d\tau^2 + e^{- 2kv\tau} \delta_{ij} dx^i dx^j.      
\ee

For $k > 0$, the static wall has a positive tension at $y=0$ and the
moving wall has a negative one at $y=R(t)$.
For $k < 0$, the situation is opposite. 
If $kv >0$, the scale factor on the moving wall is 
exponentially contracted as $\tau$ runs. This is because the wall 
is moving toward the inside the AdS space, where the scale factor decreases.  
If $kv <0$, the wall is moving toward the boundary: the scale factor increases and this is 
described as an inflation by the the observer at the domain wall. 

When we consider the region $|y|\ge R$, we have similar situation 
but with just one wall at $y=R(t)$. Exactly parallel discussion can be done. 
however this is the case which is already discussed in ref.\cite{Kraus}. 
So we so not treat it here.  

\subsection{Moving domain walls with all positive tension}
In the previous case, we imposed $\bf Z_2$ symmetry so that we 
glued two vacuum AdS solutions bounded by one static and one 
moving brane whose velocity is determined by its tension.
We also noticed that one of the brane tension must be negative.
If we want to have solution with two or more branes having the same sign,
we can not in general  impose  $\bf Z_2$ symmetry and at least 
one side of the wall must be AdS-Schwarzschild solution (AdSS).  The situation 
is just like the case of spherical shells in the Minkowski space:
Inside the shell, it is vacuum solution while it must be  Schwarzschild 
solution outside. The motion of each shell
can be is described by the Israel junction condition.
 More explicitly,
 Let there be N walls at $y=y_i(t)$ $i=1,2, \cdots, N$. Now suppose 
 that 
 the metric  in the region $y_i< y< y_{i+1}$ is described by the AdSS:
 \be
 ds^2=e^{2ky}\left[-(1-\mu_i e^{-4ky})dt^2 + d\vec{x}^2\right] 
 +\frac{dy^2}{1-\mu_i e^{-4ky}}.
 \ee 
 Then the equation of motion for the wall is given by 
 \be
 \dot{y}_i^2+1= \left(\frac{\sigma}{\sigma_c}\right)^2 
 +\frac{1}{4}(\mu_i +\mu_{i-1})e^{-4ky_i}+\frac{1}{32} \left(\frac{\sigma_c}{\sigma}\right)^2 (\mu_i 
 -\mu_{i-1})^2 e^{-8ky_i}.\label{multi}
 \ee
What is important for us at this moment  is just the existence of such solution
describing the multi-domain walls with all positive tension.

\subsection{Cosmology with Moving Domain wall}
Now let's try to realize a situation where there is a static 
domain wall at $y=y_1=0$ and there is a moving  brane at $y=y_2(t)>0$.
In the leftmost region $y<0$ we have a Ads vacuum. In between the two walls, we have
a AdSS solution with non-extremal parameter $\mu_1$. In the right most region $y>y_2(t)$,
we have another AdSS solution with parameter $\mu_2$.
The motion for the first wall is described by
 \be
 \dot{y}_1^2+1= \left(\frac{\sigma}{\sigma_c}\right)^2 
 +\frac{1}{4}( \mu_{1})e^{-4ky_1}+\frac{1}{32}\left(\frac{\sigma_c}{\sigma}\right)^2
 (\mu_1 )^2 e^{-8ky_1}.\label{twowall}
 \ee
If we want to have a static wall at $y=0$, 
 \be
 \sigma=\sigma_c\sqrt{\frac{1}{2}(1-\mu_1/4)+\frac{1}{2}\sqrt{1-\frac{5}{8}\mu_1+{1\over 16}\mu_1^2}}
   \ee
This is the condition for the static wall.

Another fact one should notice is that in the absence of the orbifold symmetry,
the effective 4 dimensional cosmological constant on the brane is not well defined. 
Decomposing the 5 dimensional Einstein 
equation to the paralell and vertical to the brane, we get 4 dimensional equation 
which contains term proportional to the induced metric  \cite{binetruy} to be identified 
as the 4 dimensional cosmological constant. It is a sum of 
bulk cosmological constant and a quadratic form of the extrinsic 
curvature:
\be
\Lambda_4=\frac{\kappa^2}{2}\Lambda_5+ 
\frac{1}{2}((K^\mu_\mu)^2-K^{\mu\nu} K_{\mu\nu}) + 
\frac{1}{4}(K_m^\sigma K_{n\sigma}g^{mn}-K_\mu^\mu K_{mn}g^{mn}),
\ee
where $K$ is the extrinsic curvature tensor and greek indices are  
5 dimensional indices while roman ones are 4 dimensional.
In the presence of the orbifold symmetry, the extrinsic curvatures 
across the brane, $K^+_{\mu\nu}$ and $K^-_{\mu\nu}$,
differ only in sign, which is indifferent in the quadratic form.  
In the absence of the symmetry, there is no way to define the effective 4 dimensional constant 
naturally. The best thing one can do is do define by hand such that the stationary
brane has zero effective cosmological constant. We will not push this idea further here. 
The motion of the second wall is described by (\ref{multi}) with two parameter $\mu_1,\mu_2$.

What happens to the stationary brane(our universe) if the other brane comes and overlaps? 
Even in the absence of the interaction between the branes, tensions will be added 
therefore we expect that the stationary brane must move and 
inflate. If both branes are infinitely thin, the impact is 
instantaneous and the consequence is expected to be small if we 
neglect the interaction of the two branes.
If the impact brane is thick, then overlapping will change the effective bulk cosmological
constant near the thin brane therefore cause a motion of the stationary brane. 
(Remember that the static condition is the balance of the bulk 
cosmological constant and the brane tension both with and without the orbifold symmetry.)
During the scattering, the thin brane moves hence inflates. 
After the thick brane pass away, the condition of the static brane is restored 
and it must stop moving.
Therefore the process of thin and thick brane scattering gives a mechanism of inflation for finite 
time. In order words, there is a natural graceful exit out of the inflation\cite{linde}.
Furthermore if the moving brane is thick enough, the scattering 
process is smooth enough so that the density fluctuation caused by the inflation may fit the present 
observation.

\section{Conclusion}

We considered moving-brane-solutions in AdS type background. In the 
first  Randall-Sundrum configuration, there are  two branes at  
fixed points of the orbifold symmetry. We proved that if 
one brane is fixed and the other brane is moving, the 
configuration is still a solution provided that the motion of the brane 
has a specific velocity  determined by its tension and the bulk cosmological constant. 
In the absence of the $\bf Z_2$ symmetry, we could construct multi-brane 
configurations patching AdS-Schwarzshild solutions. In this case, 
effective four dimensional cosmological constant on the brane is 
ambiguous. Instead, we found a condition for a brane to be stationary.
Finally, we suggest a scenario where we may have inflation on the brane universe
for finite duration of time, i.e, a graceful exit of inflation. 
Our description of the last point was very qualitative and we  did not study the 
physics of the RS1 due to the brane motion. 
We wish to comeback to  treatment of these aspects in future publication.
  
\noindent{\bf Acknowledgments}

SJS want to thank S. Nam and  P. Yi for discussion. He also want to 
express thank KIAS for its hospitality during his visit. 
This work has been supported by KOSEF 1999-2-112-003-5 and BK21.

\noindent{\bf Note added }

After typing of this manuscript was finished, we saw the 
appearance of the paper hep-th/0004206 by Horowitz, Low and Zee, 
where a wave solution were constructed in a similar fashion.

\end{document}